\documentclass[conference,letterpaper]{IEEEtran}

\usepackage[utf8]{inputenc} 
\usepackage[T1]{fontenc}
\usepackage{url}
\usepackage{ifthen}
\usepackage{cite}
\usepackage[cmex10]{amsmath} 

\usepackage{amssymb}
\usepackage[dvips]{graphicx}
\usepackage{balance}

\newtheorem{definition}{Definition}
\newtheorem{assumption}{Assumption}
\newtheorem{theorem}{Theorem}
\newtheorem{corollary}{Corollary}

\begin{document}
\title{Bayes-Optimal Convolutional AMP}


\author{
\IEEEauthorblockN{Keigo Takeuchi}
\IEEEauthorblockA{Dept.\ Electrical and Electronic Information Eng., 
Toyohashi University of Technology, 
Toyohashi 441-8580, Japan\\
Email: takeuchi@ee.tut.ac.jp} 
}


\maketitle

\begin{abstract}
To improve the convergence property of approximate message-passing (AMP), 
convolutional AMP (CAMP) has been proposed. CAMP replaces the Onsager 
correction in AMP with a convolution of messages in all preceding iterations 
while it uses the same low-complexity matched filter (MF) as AMP. This paper 
derives state evolution (SE) equations to design the Bayes-optimal denoiser 
in CAMP. Numerical results imply that CAMP 
with the Bayes-optimal denoiser---called Bayes-optimal CAMP---can achieve 
the Bayes-optimal performance for right-orthogonally invariant sensing 
matrices with low-to-moderate condition numbers.  
\end{abstract}

\section{Introduction}
\subsection{Motivation \& Contributions}
Reconstruct an unknown $N$-dimensional sparse signal vector 
$\boldsymbol{x}\in\mathbb{R}^{N}$ 
from linear and noisy $M$-dimensional measurement vector  
$\boldsymbol{y}\in\mathbb{R}^{M}$~\cite{Donoho06,Candes061}, given by  
\begin{equation} \label{model}
\boldsymbol{y} 
= \boldsymbol{A}\boldsymbol{x} + \boldsymbol{w}. 
\end{equation} 
In (\ref{model}), $\boldsymbol{A}\in\mathbb{R}^{M\times N}$ denotes a known 
sensing matrix. The additive white Gaussian noise (AWGN) vector 
$\boldsymbol{w}\in\mathbb{R}^{M}$ is composed of independent zero-mean 
Gaussian elements with variance $\sigma^{2}$. For simplicity, 
the signal vector $\boldsymbol{x}=(x_{1},\ldots,x_{N})^{\mathrm{T}}$ is assumed 
to have independent and 
identically distributed (i.i.d.) zero-mean elements with unit variance. 

A promising approach to this reconstruction issue is message-passing (MP). 
As powerful and low-complexity MP, Donoho {\it et al}.~\cite{Donoho09} 
proposed approximate message-passing (AMP). Bayes-optimal AMP can be regarded 
as an exact approximation of belief propagation (BP) in the large system 
limit~\cite{Kabashima03}, where both $M$ and $N$ tend to infinity while 
the compression rate $\delta=M/N$ is kept ${\mathcal O}(1)$. AMP was proved 
to be Bayes-optimal in a certain region of $\delta$ via state 
evolution (SE)~\cite{Bayati11,Bayati15} if the sensing matrix has zero-mean 
sub-Gaussian i.i.d.\ elements. However, AMP fails to converge for 
ill-conditioned~\cite{Rangan191} or non-zero mean~\cite{Caltagirone14} 
sensing matrices unless damping is used. 

To resolve this convergence issue, orthogonal AMP (OAMP)~\cite{Ma17} or 
equivalently vector AMP (VAMP)~\cite{Rangan192} were proposed. A prototype 
of OAMP/VAMP was presented in a pioneering paper~\cite[Appendix~D]{Opper05}.  
Bayes-optimal OAMP/VAMP can be regarded as an exact 
approximation~\cite{Cespedes14,Takeuchi201} of 
expectation propagation (EP)~\cite{Minka01} in the large system 
limit. OAMP/VAMP was proved to be Bayes-optimal 
in a certain region of $\delta$ via SE~\cite{Rangan192,Takeuchi201} if 
the sensing matrix is right-orthogonally invariant. 

A disadvantage of Bayes-optimal OAMP/VAMP is use of the linear minimum 
mean-square error (LMMSE) filter. While the singular-value decomposition (SVD) 
of the sensing matrix circumvents per-iteration computation of the LMMSE 
filter~\cite{Rangan192}, the SVD itself is high complexity unless the sensing 
matrix has a special structure. 

This paper proposes novel MP to solve the convergence issue in AMP and 
the high-complexity issue in OAMP/VAMP. The main idea is a generalization 
of conventional MP to long-memory MP (LM-MP)~\cite{Takeuchi19}. LM-MP exploits 
messages in all preceding iterations for each message update while conventional 
MP only utilizes them in the latest iteration. LM-MP aims to improve the 
convergence property of AMP ultimately up to that of OAMP/VAMP, without 
using the LMMSE filter. 

As a solvable instance of LM-MP, convolutional AMP (CAMP) has been proposed 
in \cite{Takeuchi202}. CAMP replaces the so-called Onsager 
correction in AMP with a convolution of messages in all preceding iterations 
while it uses the same low-complexity matched filter (MF) as AMP. 
Tap coefficients in the convolution are designed so as to guarantee asymptotic 
Gaussianity of estimation errors for all right-orthogonally invariant sensing 
matrices. However, Bayes-optimal denoisers were not designed in CAMP 
because \cite{Takeuchi202} presented no SE analysis to evaluate the mean-square 
errors (MSEs) before/after denoising.  

The main contribution of this paper is the SE analysis to establish CAMP with 
Bayes-optimal denoisers. Two-dimensional difference equations---called SE 
equations---are derived to describe the dynamics of the MSEs before/after 
denoising. The SE equations are utilized to compute variance parameters  
that are used in the Bayes-optimal denoisers. CAMP with Bayes-optimal denoisers 
is proved to achieve the same performance as OAMP/VAMP if the SE equations 
converge. Thus, CAMP with Bayes-optimal denoisers is called 
Bayes-optimal CAMP.   

\subsection{Related Works}
As related works, a similar approach to LM-MP~\cite{Takeuchi19} was 
considered via non-rigorous dynamical functional theory~\cite{Opper16} 
and rigorous SE~\cite{Fan20}. LM-MP~\cite{Takeuchi19} imposes the 
orthogonality between estimation errors before/after denoising  
as considered in OAMP~\cite{Ma17}, while \cite{Opper16,Fan20} requires no 
restriction. The orthogonality enables a systematic design of LM-MP that 
satisfies asymptotic Gaussianity of estimation errors. 
 
As another related work, memory AMP (MAMP)~\cite{Liu20} was proposed 
after submission of a long version~\cite{Takeuchi203} for this paper. 
MAMP utilizes all preceding messages in damping before/after 
denoising while CAMP exploits them in the Onsager correction. The LM-MP 
framework~\cite{Takeuchi19} was used to design the LM damping in MAMP. 

\section{Convolutional Approximate Message-Passing}
CAMP~\cite{Takeuchi202} computes an estimator 
$\boldsymbol{x}_{t}\in\mathbb{R}^{N}$ of the signal vector 
$\boldsymbol{x}$ in iteration~$t$ from the information about the 
measurement vector $\boldsymbol{y}$ and the sensing matrix $\boldsymbol{A}$ in 
(\ref{model}), 
\begin{equation} \label{denoising}
\boldsymbol{x}_{t+1} 
= f_{t}(\boldsymbol{x}_{t} + \boldsymbol{A}^{\mathrm{T}}\boldsymbol{z}_{t}), 
\end{equation}
with $\boldsymbol{x}_{0}=\boldsymbol{0}$ and  
$\boldsymbol{z}_{0}=\boldsymbol{y}$. For $t>0$,  
\begin{equation} \label{z}
\boldsymbol{z}_{t} 
= \boldsymbol{y} - \boldsymbol{A}\boldsymbol{x}_{t} 
+ \sum_{\tau=0}^{t-1}\xi_{\tau}^{(t-1)}\left(
 \theta_{t-\tau}\boldsymbol{A}\boldsymbol{A}^{\mathrm{T}} 
 - g_{t-\tau}\boldsymbol{I}_{M}
\right)\boldsymbol{z}_{\tau}. 
\end{equation}

In (\ref{denoising}), a Lipschitz-continuous scalar denoiser 
$f_{t}:\mathbb{R}\to\mathbb{R}$ is applied element-wisely. 
In (\ref{z}), the notation $\xi_{t'}^{(t)}
=\prod_{\tau=t'}^{t}\xi_{\tau}$ for $t'\leq t$ is defined via 
\begin{equation} \label{xi}
\xi_{t} = \left\langle
 f_{t}'(\boldsymbol{x}_{t} + \boldsymbol{A}^{\mathrm{T}}\boldsymbol{z}_{t})
\right\rangle, 
\end{equation}
where $\langle \boldsymbol{v}\rangle=
N^{-1}\sum_{n=1}^{N}v_{n}$ denotes the arithmetic 
average of $\boldsymbol{v}=(v_{1},\ldots,v_{N})^{\mathrm{T}}$. 

The last term on the right-hand side (RHS) in (\ref{z}) is called the Onsager 
correction, which is the convolution of the messages $\{\boldsymbol{z}_{0},
\ldots,\boldsymbol{z}_{t-1}\}$ in all preceding iterations. The tap 
coefficients $\{g_{t}\}$ are designed via the LM-MP 
framework~\cite{Takeuchi19} so as to guarantee asymptotic Gaussianity 
of the estimation error $\boldsymbol{x}_{t}-\boldsymbol{x}$.  
The other tap coefficients $\{\theta_{t}\}$ have been introduced in this paper 
to improve the convergence property of CAMP. Note that $\theta_{t}=0$ for 
any $t\geq1$ was used in the original CAMP~\cite{Takeuchi202}.  

Once asymptotic Gaussianity is established, the Bayes-optimal denoiser 
$f_{t}(\boldsymbol{u}_{t}) = \mathbb{E}[\boldsymbol{x} | \boldsymbol{u}_{t}]$ 
can be designed via the virtual AWGN observation, 
\begin{equation} \label{AWGN}
\boldsymbol{u}_{t}
=\boldsymbol{x}+\boldsymbol{\omega}_{t},\quad  
\boldsymbol{\omega}_{t}\sim\mathcal{N}(\boldsymbol{0},a_{t,t}
\boldsymbol{I}_{N}).
\end{equation} 
Since $\boldsymbol{x}$ has been assumed to have i.i.d.\ elements, 
the Bayes-optimal denoiser is separable. 
The purpose of this paper is to evaluate the variance $a_{t,t}$ in the AWGN 
observation~(\ref{AWGN}) via SE. 

\section{Design}
\subsection{Asymptotic Gaussianity}
The tap coefficients $\{g_{t}\}$ in (\ref{z}) are determined so as to 
realize the asymptotic Gaussianity of the estimation error 
$\boldsymbol{h}_{t}= \boldsymbol{x}_{t} 
+ \boldsymbol{A}^{\mathrm{T}}\boldsymbol{z}_{t} - \boldsymbol{x}$ before 
denoising in (\ref{denoising}), i.e.\ almost surely 
\begin{IEEEeqnarray}{rl}
&\lim_{M=\delta N\to\infty}\frac{1}{N}\left\{
 f_{t}(\boldsymbol{x} + \boldsymbol{h}_{t}) - \boldsymbol{x}
\right\}^{\mathrm{T}}
\left\{
 f_{t'}(\boldsymbol{x} + \boldsymbol{h}_{t'}) - \boldsymbol{x}
\right\}  \nonumber \\
=& \mathbb{E}\left[
 \{f_{t}(x_{1} + h_{t}) - x_{1}\}\{f_{t'}(x_{1} + h_{t'}) - x_{1}\}
\right]\equiv d_{t+1,t'+1},  \nonumber \\
\label{Gaussianity}
\end{IEEEeqnarray}
where $(h_{t}, h_{t'})^{\mathrm{T}}\sim\mathcal{N}(\boldsymbol{0},
\boldsymbol{\Sigma})$ is an independent zero-mean Gaussian vector 
with covariance 
\begin{equation}
\boldsymbol{\Sigma} 
= \begin{bmatrix}
a_{t,t} & a_{t,t'} \\ 
a_{t',t} & a_{t',t'} 
\end{bmatrix},  
\end{equation}
with 
\begin{equation} \label{a_tt} 
a_{t,t'}\equiv\lim_{M=\delta N\to\infty}\frac{1}{N}
\mathbb{E}[\boldsymbol{h}_{t}^{\mathrm{T}}\boldsymbol{h}_{t'}].
\end{equation}
The asymptotic Gaussianity~(\ref{Gaussianity}) implies that the estimation 
errors $\boldsymbol{h}_{t}$ and $\boldsymbol{h}_{t'}$ can be treated as if 
they followed the zero-mean Gaussian distribution with 
$\mathbb{E}[\boldsymbol{h}_{t}\boldsymbol{h}_{t'}^{\mathrm{T}}]
=a_{t,t'}\boldsymbol{I}_{N}$, as long as the covariance on 
the left-hand side (LHS) of (\ref{Gaussianity}) is considered.   

The asymptotic Gaussianity is proved via a unified framework of 
SE~\cite{Takeuchi19}, which proposed a general error model and proved 
the asymptotic Gaussianity of the estimation error before denoising in the 
general error model. Thus, it is sufficient to prove that the general error 
model in \cite{Takeuchi19} contains the error model of CAMP.  

An important assumption in the general error model is orthogonal invariance 
of the sensing matrix $\boldsymbol{A}$. 
\begin{definition}
An orthogonal matrix $\boldsymbol{V}$ is said to be 
Haar-distributed if $\boldsymbol{V}$ is orthogonally invariant, 
i.e.\ $\boldsymbol{V}\sim\boldsymbol{\Phi}\boldsymbol{V}\boldsymbol{\Psi}$ 
for all orthogonal matrices $\boldsymbol{\Phi}, 
\boldsymbol{\Psi}$ independent of $\boldsymbol{V}$. 
\end{definition}
\begin{assumption} \label{assumption_A} 
The sensing matrix $\boldsymbol{A}$ is right-orthogonally invariant, i.e.\ 
$\boldsymbol{A}\sim\boldsymbol{A}\boldsymbol{\Psi}$ for any orthogonal 
matrix $\boldsymbol{\Psi}$ independent of $\boldsymbol{A}$. 
More precisely, the $N\times N$ orthogonal matrix $\boldsymbol{V}$ 
in the SVD $\boldsymbol{A}=\boldsymbol{U}\boldsymbol{\Sigma}
\boldsymbol{V}^{\mathrm{T}}$ is Haar-distributed and independent of 
$\boldsymbol{U}\boldsymbol{\Sigma}$. Furthermore, the empirical eigenvalue 
distribution of $\boldsymbol{A}^{\mathrm{T}}\boldsymbol{A}$ converges almost 
surely to a compactly supported deterministic distribution with unit first 
moment in the large system limit. 
\end{assumption}

Throughout this paper, Assumption~\ref{assumption_A} is postulated. 
Assumption~\ref{assumption_A} holds when $\boldsymbol{A}$ has zero-mean 
i.i.d.\ Gaussian elements with variance $1/M$. 
The asymptotic Gaussianity depends heavily on the 
Haar assumption of $\boldsymbol{V}$. The Haar orthogonal transform 
$\boldsymbol{V}\boldsymbol{a}$ of any vector $\boldsymbol{a}\in\mathbb{R}^{N}$ 
is distributed as $N^{-1/2}\|\boldsymbol{a}\|\boldsymbol{z}$ in which 
$\boldsymbol{z}\sim\mathcal{N}(\boldsymbol{0},\boldsymbol{I}_{N})$ is a 
standard Gaussian vector and independent of $\|\boldsymbol{a}\|$. When  
the amplitude $N^{-1/2}\|\boldsymbol{a}\|$ tends to a constant as $N\to\infty$, 
the vector $\boldsymbol{V}\boldsymbol{a}$ looks like a Gaussian vector. 
This is a rough intuition on the asymptotic Gaussianity. 

To present a closed-form of $\{g_{t}\}$ realizing the asymptotic Gaussianity, 
we define the $\eta$-transform of the asymptotic eigenvalue distribution 
of $\boldsymbol{A}^{\mathrm{T}}\boldsymbol{A}$~\cite{Tulino04} as 
\begin{equation}
\eta(x) 
= \lim_{M=\delta N\to\infty}\frac{1}{N}\mathrm{Tr}\left\{
 \left(
  \boldsymbol{I}_{N} + x\boldsymbol{A}^{\mathrm{T}}\boldsymbol{A}
 \right)^{-1}
\right\}. 
\end{equation}
Furthermore, let $G(z)$ denote the generating function 
of the tap coefficients $\{g_{t}\}$, 
\begin{equation} \label{G}
G(z) = \sum_{t=0}^{\infty}g_{t}z^{-t},
\quad g_{0}=1. 
\end{equation}
Similarly, the generating function $\Theta(z)$ of $\{\theta_{t}\}$ is defined 
in the same manner, with $\theta_{0}=1$. The tap coefficients $\{g_{t}\}$ 
are given via the generating function $G(z)$. 
\begin{theorem} \label{theorem1}
For a fixed generating function $\Theta(z)$, suppose that the generating 
function $G(z)$ of $\{g_{t}\}$ satisfies  
\begin{equation} \label{tap_coefficient_g}
\eta\left(
 \frac{1 - (1-z^{-1})\Theta(z)}{(1-z^{-1})G(z)}
\right) = (1 - z^{-1})\Theta(z). 
\end{equation}
Then, the asymptotic Gaussianity~(\ref{Gaussianity}) holds.  
\end{theorem}
\begin{IEEEproof}
See \cite[Theorems~1, 2, and 3]{Takeuchi203}. 
\end{IEEEproof}

The proof of Theorem~\ref{theorem1} is essentially the same as in the 
original CAMP~\cite{Takeuchi202} with $\theta_{0}=1$ and 
$\theta_{t}=0$ for $t>0$. For special sensing matrices $\boldsymbol{A}$, 
the tap coefficients $\{g_{t}\}$ are explicitly given as follows: 
\begin{corollary} \label{corollary1}
Suppose that the sensing matrix $\boldsymbol{A}$ has independent Gaussian 
elements with mean $\sqrt{\gamma/M}$ and variance $(1-\gamma)/M$ for any 
$\gamma\in[0,1)$. Then, (\ref{tap_coefficient_g}) reduces to 
\begin{equation}
g_{t} = \left(
 1 - \frac{1}{\delta}
\right)\theta_{t} 
+ \frac{1}{\delta}\sum_{\tau=0}^{t}(\theta_{\tau} - \theta_{\tau-1})\theta_{t-\tau}. 
\end{equation}
\end{corollary}
\begin{IEEEproof}
See \cite[Corollary~1]{Takeuchi203}. 
\end{IEEEproof}

Corollary~\ref{corollary1} implies that CAMP reduces to 
AMP for $\theta_{0}=1$ and $\theta_{t}=0$ for $t>0$. Thus, CAMP 
has no ability to resolve the non-zero mean case~\cite{Caltagirone14}. 

\begin{corollary}
Suppose that the sensing matrix $\boldsymbol{A}$ has $M$ identical singular 
values for $M\leq N$, i.e.\ $\boldsymbol{A}\boldsymbol{A}^{\mathrm{T}}
=\delta^{-1}\boldsymbol{I}_{M}$. Then, (\ref{tap_coefficient_g}) with 
$\theta_{t}=0$ for $t>0$ reduces to $g_{t}=1-\delta^{-1}$ for all $t>0$.  
\end{corollary}
\begin{IEEEproof}
See \cite[Corollary~2]{Takeuchi203}. 
\end{IEEEproof}

Note that $\theta_{t}=0$ can be set for $t>0$, without loss of generality, 
since $\theta_{t-\tau}\boldsymbol{A}\boldsymbol{A}^{\mathrm{T}} 
- g_{t-\tau}\boldsymbol{I}_{M} = (\delta^{-1}\theta_{t-\tau} - g_{t-\tau})
\boldsymbol{I}_{M}$ holds on the Onsager term in (\ref{z}). 

To present the non-identical singular-value case, we define the convolution 
operator $*$ as 
\begin{equation} \label{convolution}
a_{t+i}*b_{t+j} 
= \sum_{\tau=0}^{t}a_{\tau+i}b_{t-\tau+j}
\end{equation}
for two sequences $\{a_{\tau}, b_{\tau}\}_{\tau=0}^{\infty}$, in which  
$a_{\tau}=0$ and $b_{\tau}=0$ are assumed for all $\tau<0$. 

\begin{corollary} \label{corollary3}
Suppose that the sensing matrix $\boldsymbol{A}$ has non-zero singular 
values $\sigma_{0}\geq\cdots\geq\sigma_{M-1}>0$ satisfying condition number 
$\kappa=\sigma_{0}/\sigma_{M-1}>1$, $\sigma_{m}/\sigma_{m-1}=\kappa^{-1/(M-1)}$, 
and $\sigma_{0}^{2}=N(1-\kappa^{-2/(M-1)})/(1-\kappa^{-2M/(M-1)})$, and that 
there is some $t_{1}\in\mathbb{N}$ such that $\theta_{t}=0$ holds for all 
$t>t_{1}$. Let $\alpha_{0}^{(j)}=1$ and 
\begin{equation}
\alpha_{t}^{(j)} 
= \left\{
\begin{array}{cl}
\frac{C^{t/j}}{(t/j)!}\bar{\theta}_{j}^{t/j} & 
\hbox{if $t$ is divisible by $j$,} \\ 
0 & \hbox{otherwise} 
\end{array}
\right.
\end{equation}
for $t\in\mathbb{N}$ and $j\in\{1,\ldots,t_{1}\}$, 
with $\bar{\theta}_{t}=\theta_{t-1}-\theta_{t}$ and  
$C=2\delta^{-1}\ln\kappa$. 
Define $p_{0}=\bar{q}_{0}=1$ and 
\begin{equation}
p_{t}
= - \frac{\beta_{t}^{(t_{1})}}{\kappa^{2}-1},  
\end{equation}
\begin{equation}
\bar{q}_{t} 
= \frac{1}{\bar{\theta}_{1}}\left(
 \frac{\beta_{t+1}^{(t_{1})}}{C} 
 - \sum_{\tau=1}^{t_{1}}\bar{\theta}_{\tau+1}\bar{q}_{t-\tau} 
\right)
\end{equation}
for $t>0$, 
with $\beta_{t}^{(t_{1})}=\alpha_{t}^{(1)}*\alpha_{t}^{(2)}*\cdots*
\alpha_{t}^{(t_{1})}$. 
Then, (\ref{tap_coefficient_g}) reduces to 
\begin{equation} \label{g_CAMP}
g_{t} = p_{t} - \sum_{\tau=1}^{t}q_{\tau}g_{t-\tau}, 
\end{equation}
with 
\begin{equation}
q_{t} = \bar{q}_{t} - \bar{q}_{t-1}. 
\end{equation}
\end{corollary}
\begin{IEEEproof}
See \cite[Corollary~3]{Takeuchi203}. 
\end{IEEEproof}
Note that $G(z)=P(z)/Q(z)$ holds, with $P(z)$ and $Q(z)$ denoting the 
generating functions of $\{p_{t}\}$ and $\{q_{t}\}$ in 
Corollary~\ref{corollary3}, respectively. 

\subsection{State Evolution}
We have so far designed the tap coefficients $\{g_{t}\}$ that realize 
the asymptotic Gaussianity~(\ref{Gaussianity}) for fixed tap coefficients 
$\{\theta_{t}\}$. We next evaluate the asymptotic MSEs $\{a_{t,t}\}$ before 
denoising in (\ref{a_tt}). 

We use SE to derive SE equations that describe the dynamics of $\{a_{t,t}\}$. 
The derived SE equations are two-dimensional difference equations with 
respect to the covariance parameters $\{a_{t,t'}\}$ and $\{d_{t,t'}\}$ before and 
after denoising in (\ref{a_tt}) and (\ref{Gaussianity}), respectively. 

To present the SE equations, we introduce several notations. 
In terms of their implementations, rather than the proof, 
we re-write the generating function~(\ref{G}) as   
\begin{equation}
G(z)=\frac{\sum_{t=0}^{\infty}p_{t}z^{-t}}{\sum_{t=0}^{\infty}q_{t}z^{-t}},
\end{equation} 
with $p_{0}=1$, $q_{0}=1$. In particular, $g_{t}=p_{t}$ holds when  
$q_{t}=0$ is selected for all $t>0$. 

We define $\bar{\xi}_{t}$ as an asymptotic variable associated 
with $\xi_{t}$ given in (\ref{xi}), 
\begin{equation} \label{xi_bar}
\bar{\xi}_{t} 
= \mathbb{E}[f_{t}'(x_{1}+h_{t})].
\end{equation}
In (\ref{xi_bar}), $h_{t}$ is an independent zero-mean Gaussian random 
variable with variance $a_{t,t}$. Thus, $\bar{\xi}_{t}$ is a function of 
$a_{t,t}$. The notation $\bar{\xi}_{t'}^{(t)}=\prod_{\tau=t'}^{t}\bar{\xi}_{\tau}$ 
is defined in the same manner as in $\xi_{t'}^{(t)}$. Furthermore, we use 
the notational convention $\bar{\xi}_{t'}^{(t)}=1$ for $t'>t$. 

\begin{theorem} \label{theorem2}
Suppose that the generating functions of the tap coefficients $\{g_{t}\}$ 
and $\{\theta_{t}\}$ satisfy the condition~(\ref{tap_coefficient_g}) in 
Theorem~\ref{theorem1}.  
Define $r_{t}=q_{t}*\theta_{t}$ via the convolution~(\ref{convolution}) and   
\begin{IEEEeqnarray}{rl}
\mathfrak{D}_{\tau',\tau}
&= (p_{\tau'+\tau} - p_{\tau'+\tau+1})*q_{\tau} + (p_{\tau} - p_{\tau-1})*q_{\tau'+\tau+1} 
\nonumber \\
+& (p_{\tau-1}-p_{\tau})*r_{\tau'+\tau+1} + (r_{\tau} - r_{\tau-1})*p_{\tau'+\tau+1}
 \nonumber \\
+& p_{\tau}*(r_{\tau'+\tau} - \delta_{\tau',0}r_{\tau}) 
- r_{\tau}*(p_{\tau'+\tau} - \delta_{\tau',0}p_{\tau}), \label{D}
\end{IEEEeqnarray}
where $\delta_{t,t'}$ denotes the Kronecker delta. 
Then, the covariance parameters $\{a_{t,t'}\}$ in (\ref{a_tt}) satisfy 
\begin{IEEEeqnarray}{rl}
\sum_{\tau'=0}^{t'}\sum_{\tau=0}^{t}\bar{\xi}_{t'-\tau'}^{(t'-1)}\bar{\xi}_{t-\tau}^{(t-1)}
\Big\{ \mathfrak{D}_{\tau',\tau}a_{t'-\tau',t-\tau}&  
\nonumber \\
- (p_{\tau}*r_{\tau'+\tau+1} - r_{\tau}*p_{\tau'+\tau+1})d_{t'-\tau',t-\tau}&
\nonumber \\
- \sigma^{2}\left[
 (q_{\tau'}q_{\tau})*(\theta_{\tau'+\tau} - \theta_{\tau'+\tau+1})
\right]&\Big\} 
=0, \label{SE_equation}  
\end{IEEEeqnarray}
with 
\begin{IEEEeqnarray}{rl}
&(q_{t'}q_{t})*(\theta_{t'+t} - \theta_{t'+t+1}) \nonumber \\
=& \sum_{\tau'=0}^{t'}\sum_{\tau=0}^{t}q_{\tau'}q_{\tau}
(\theta_{t'-\tau'+t-\tau} - \theta_{t'-\tau'+t-\tau+1}). 
\end{IEEEeqnarray}
In these expressions, all variables with negative indices are set to zero. 
The covariance parameters $\{d_{t,t'}\}$ are given in (\ref{Gaussianity}) 
with notational convention $f_{-1}(\cdot)=0$. 
\end{theorem}
\begin{IEEEproof}
See \cite[Theorem~4]{Takeuchi203}. 
\end{IEEEproof}

The coupled SE equations~(\ref{Gaussianity}) and (\ref{SE_equation}) 
describe the dynamics of the covariance parameters $\{a_{t,t'}\}$ 
and $\{d_{t,t'}\}$. The SE equations can be solved with the initial 
condition $d_{0,0}=N^{-1}\mathbb{E}[\|\boldsymbol{x}\|^{2}]=1$ and 
the boundary conditions $a_{t,t'}=d_{t,t'}=0$ for $t<0$ or $t'<0$. 

We next optimize the denoiser $f_{t}$ to establish Bayes-optimal CAMP. 
The asymptotic Gaussianity~(\ref{Gaussianity}) indicates that 
the posterior mean estimator $f_{t}(\boldsymbol{u}_{t})
=\mathbb{E}[\boldsymbol{x}|\boldsymbol{u}_{t}]\equiv 
f_{\mathrm{opt}}(\boldsymbol{u}_{t}; a_{t,t})$ 
of $\boldsymbol{x}$ 
given the AWGN observation~(\ref{AWGN}) minimizes the asymptotic 
MSE~$d_{t+1,t+1}$ in (\ref{Gaussianity}) after denoising in CAMP. Thus, 
we use the posterior mean estimator as the Bayes-optimal denoiser.

\begin{theorem} \label{theorem3}
Use the Bayes-optimal denoiser. Suppose that the SE 
equations~(\ref{Gaussianity}) and (\ref{SE_equation}) converge, 
i.e.\ $\lim_{t',t\to\infty}a_{t',t}=a_{\mathrm{s}}$ and 
$\lim_{t',t\to\infty}d_{t',t}=d_{\mathrm{s}}$. If $\Theta(\xi_{\mathrm{s}}^{-1})=1$ 
and $1+(\xi_{\mathrm{s}}-1)d\Theta(\xi_{\mathrm{s}}^{-1})/(dz^{-1})\neq0$ 
hold for $\xi_{\mathrm{s}}=d_{\mathrm{s}}/a_{\mathrm{s}}$, then 
the fixed-point (FP) $(a_{\mathrm{s}}, d_{\mathrm{s}})$ satisfies 
\begin{equation} \label{fixed_point} 
a_{\mathrm{s}}
= \frac{\sigma^{2}}
{R_{\boldsymbol{A}^{\mathrm{T}}\boldsymbol{A}}(-d_{\mathrm{s}}/\sigma^{2})}, 
\;
d_{\mathrm{s}}=\mathbb{E}\left[
 \{f_{\mathrm{opt}}(x_{1}+h_{\mathrm{s}}; a_{\mathrm{s}}) - x_{1}\}^{2}
\right],
\end{equation}
with $h_{\mathrm{s}}\sim\mathcal{N}(0,a_{\mathrm{s}})$, 
where $R_{\boldsymbol{A}^{\mathrm{T}}\boldsymbol{A}}(x)$ denotes the R-transform of the 
asymptotic eigenvalue distribution 
of $\boldsymbol{A}^{\mathrm{T}}\boldsymbol{A}$~\cite{Tulino04}. 
\end{theorem}
\begin{IEEEproof}
Without loss of generality, we assume $p_{t}=g_{t}$ and $q_{t}=\delta_{t,0}$. 
Since the SE equations have been assumed to converge, (\ref{SE_equation}) 
reduces to 
\begin{IEEEeqnarray}{rl}
\sum_{\tau'=0}^{\infty}\sum_{\tau=0}^{\infty}\xi_{\mathrm{s}}^{\tau'+\tau}
&\Big\{ \mathfrak{D}_{\tau',\tau}a_{\mathrm{s}}  
- (g_{\tau}*\theta_{\tau'+\tau+1} - \theta_{\tau}*g_{\tau'+\tau+1})d_{\mathrm{s}}
\nonumber \\
&- \sigma^{2}(\theta_{\tau'+\tau} - \theta_{\tau'+\tau+1})
\Big\} 
=0, \label{SE_equation_FP} 
\end{IEEEeqnarray}
as $t, t'\to\infty$, with 
\begin{IEEEeqnarray}{rl}
\mathfrak{D}_{\tau',\tau}
=& g_{\tau'+\tau} - g_{\tau'+\tau+1} +(g_{\tau-1}-g_{\tau})*\theta_{\tau'+\tau+1}  
\nonumber \\
&+ (\theta_{\tau} - \theta_{\tau-1})*g_{\tau'+\tau+1}
+ g_{\tau}*(\theta_{\tau'+\tau} - \delta_{\tau',0}\theta_{\tau}) 
 \nonumber \\
&- \theta_{\tau}*(g_{\tau'+\tau} - \delta_{\tau',0}g_{\tau}). \label{D_tmp}
\end{IEEEeqnarray}

We next re-write (\ref{SE_equation_FP}) via the Z-transform. 
Define the Z-transform of a two-dimensional array 
$\{h_{t',t}\}_{t',t=0}^{\infty}$ as 
\begin{equation}
H(y,z) = \sum_{t',t=0}^{\infty}h_{t',t}y^{-t'}z^{-t}. 
\end{equation}
It is an elementary exercise to confirm that the Z-transform of 
$\{\mathcal{D}_{t',t}\}$ given in (\ref{D_tmp}) is equal 
to~\cite{Takeuchi203}   
\begin{IEEEeqnarray}{rl}
F_{G,\Theta}(y,z) 
=& (y^{-1}+z^{-1}-1)[G(z)\Delta_{\Theta} - \Theta(z)\Delta_{G}]
\nonumber \\
&+ \Delta_{G_{1}} - \Delta_{G}, \label{F_G}
\end{IEEEeqnarray}
where $G(z)$ and $\Theta(z)$ are the Z-transforms of 
$\{g_{t}\}$ and $\{\theta_{t}\}$, respectively, with
\begin{equation}
F_{1}(z)=z^{-1}F(z),\quad
\Delta_{F(y)} 
= \frac{F(y) - F(z)}{y^{-1}-z^{-1}}. 
\end{equation}
Representing the remaining terms in (\ref{SE_equation_FP}) with the 
Z-transform, we obtain  
\begin{equation} 
F_{G,\Theta}(y,z)\frac{d_{\mathrm{s}}}{\xi_{\mathrm{s}}}
=\{G(z)\Delta_{\Theta} - \Theta(z)\Delta_{G}\}d_{\mathrm{s}}  
+ (\Delta_{\Theta_{1}} - \Delta_{\Theta})\sigma^{2}  
\label{SE_equation_FP_Z}
\end{equation}
in the limit $y, z\to\xi_{\mathrm{s}}^{-1}$, where we have used 
the identity $\xi_{\mathrm{s}}=d_{\mathrm{s}}/a_{\mathrm{s}}$ for the 
Bayes-optimal denoiser.  

We simplify (\ref{SE_equation_FP_Z}) via series-expansion. 
Series-expanding $\Delta_{F}$ and $\Delta_{F_{1}}$ with respect to $z^{-1}$ 
at $z=y$ up to the first order, we have 
\begin{equation}
\left\{
 1 + (\xi_{\mathrm{s}}-1)
 \frac{d\Theta}{dz^{-1}}(\xi_{\mathrm{s}}^{-1}) 
\right\}\left\{
 \frac{G(\xi_{\mathrm{s}}^{-1})d_{\mathrm{s}}}{\xi_{\mathrm{s}}}
 - \sigma^{2}
\right\} 
= 0
\end{equation}
under the assumptions of $\Theta(\xi_{\mathrm{s}}^{-1})=1$. 
Since $1+(\xi_{\mathrm{s}}-1)d\Theta(\xi_{\mathrm{s}}^{-1})/(dz^{-1})\neq0$ 
has been assumed, we arrive at   
\begin{equation} \label{G_idenity}
\frac{G(\xi_{\mathrm{s}}^{-1})}{\xi_{\mathrm{s}}} 
= \frac{\sigma^{2}}{d_{\mathrm{s}}}.  
\end{equation}

To prove the FP~(\ref{fixed_point}), we utilize the following 
relationship between the $\eta$-transform and the 
R-transform~\cite[Eq.~(2.74)]{Tulino04}:  
\begin{equation} \label{relationship} 
\eta(x) = \frac{1}{1 + xR_{\boldsymbol{A}^{\mathrm{T}}\boldsymbol{A}}(-x\eta(x))}. 
\end{equation}
Using (\ref{tap_coefficient_g}) to evaluate (\ref{relationship}) at 
$x=x^{*}$ given by  
\begin{equation} \label{pole}
x^{*} = \frac{1-(1-z^{-1})\Theta(z)}{(1-z^{-1})G(z)}, 
\end{equation}
we obtain 
\begin{equation}
G(z) 
= \Theta(z) R_{\boldsymbol{A}^{\mathrm{T}}\boldsymbol{A}}\left(
 -\frac{1 - (1-z^{-1})\Theta(z)}{G(z)}\Theta(z)
\right).
\end{equation}
Letting $z=\xi_{\mathrm{s}}^{-1}$ and applying the assumption 
$\Theta(\xi_{\mathrm{s}}^{-1})=1$ yield  
\begin{equation}
G(\xi_{\mathrm{s}}^{-1}) 
= R_{\boldsymbol{A}^{\mathrm{T}}\boldsymbol{A}}\left(
 - \frac{\xi_{\mathrm{s}}}{G(\xi_{\mathrm{s}}^{-1})} 
\right). 
\end{equation}
Substituting (\ref{G_idenity}) into this identity and using 
$\xi_{\mathrm{s}}=d_{\mathrm{s}}/a_{\mathrm{s}}$, we arrive at the 
FP~(\ref{fixed_point}). 
\end{IEEEproof}

The FP~(\ref{fixed_point}) is equal to that of the Bayes-optimal 
performance~\cite{Takeda06,Tulino13,Barbier18}. 
This implies that CAMP with the Bayes-optimal denoiser is Bayes-optimal 
if it converges as $t\to\infty$ and if the FP~(\ref{fixed_point}) is unique. 
Thus, we refer to CAMP with the Bayes-optimal denoiser as Bayes-optimal 
CAMP or simply as CAMP. 

The original CAMP~\cite{Takeuchi202} with $\theta_{0}=1$ and $\theta_{t}=0$ 
for $t>0$ satisfies the conditions with respect to $\Theta(z)$ in 
Theorem~\ref{theorem3}. Thus, the parameters $\{\theta_{t}\}$ only 
contribute to the convergence properties of CAMP. Numerical evaluation in 
the next section implies that using non-zero $\theta_{t}$ improves the 
stability of CAMP.

\section{Numerical Results}
In all numerical results, we assume the Bernoulli-Gaussian (BG) prior with 
signal density $\rho\in[0, 1]$. Each signal $x_{n}$ takes zero with 
probability $1-\rho$. Otherwise, $x_{n}$ is sampled from the zero-mean 
Gaussian distribution with variance $1/\rho$. 

As a practical alternative of $\boldsymbol{V}$ in 
Assumption~\ref{assumption_A}, we used Hadamard matrices with random  
permutation in numerical simulations. 
See Corollary~\ref{corollary3} for the details of singular values. 

For simplicity, we assume $\theta_{t}=0$ for all $t>2$. 
To impose the condition $\Theta(a_{\mathrm{s}}/d_{\mathrm{s}})=1$ in 
Theorem~\ref{theorem3}, we use $\theta_{0}=1$, 
$\theta_{1}=-\theta d_{\mathrm{s}}/a_{\mathrm{s}}$, and 
$\theta_{2}=\theta\in\mathbb{R}$, in which 
$(a_{\mathrm{s}}, d_{\mathrm{s}})$ is a solution to the FP 
equations~(\ref{fixed_point}). In particular, the CAMP reduces to the 
original CAMP in \cite{Takeuchi202} for $\theta=0$.  

\begin{figure}[t]
\begin{center}
\includegraphics[width=\hsize]{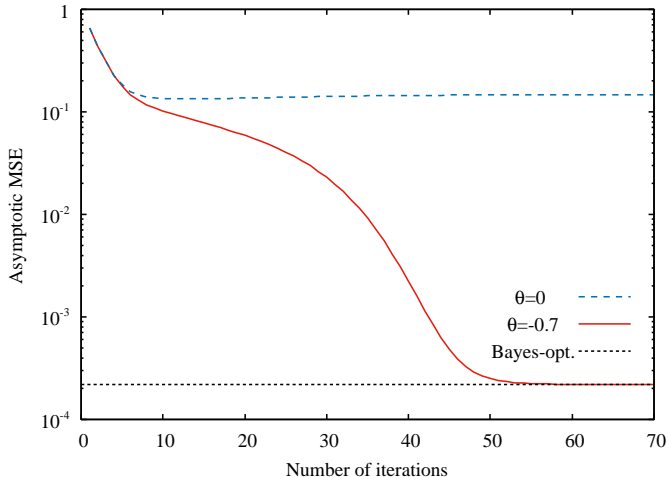}
\caption{
Asymptotic MSE $d_{t+1,t+1}$ versus the number of iterations~$t$ for the CAMP. 
$\delta=0.5$, $\rho=0.1$, condition number~$\kappa=17$, 
and $1/\sigma^{2}=30$~dB.  
}
\label{fig1} 
\end{center}
\end{figure}

We first solve the SE equations~(\ref{Gaussianity}) and (\ref{SE_equation}) 
to investigate the convergence property of CAMP. 
Figure~\ref{fig1} shows the asymptotic MSEs of the original 
CAMP with $\theta=0$~\cite{Takeuchi202} and of the proposed CAMP with 
$\theta=-0.7$. The proposed CAMP can achieve the Bayes-optimal 
MSE~\cite{Takeda06,Tulino13,Barbier18} while the original CAMP with 
$\theta=0$ fails to converge. This implies that use of non-zero $\theta\neq0$ 
improves the stability of CAMP. 
 
\begin{figure}[t]
\begin{center}
\includegraphics[width=\hsize]{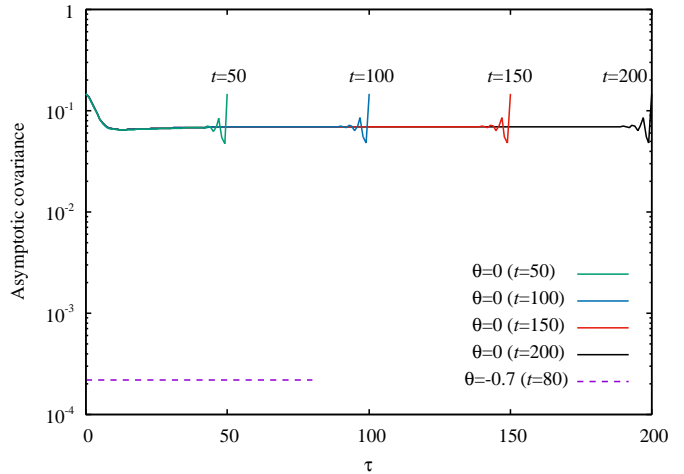}
\caption{
Asymptotic covariance $d_{\tau, t}$ versus $\tau$ for the CAMP. $\delta=0.5$, 
$\rho=0.1$, condition number~$\kappa=17$, and $1/\sigma^{2}=30$~dB.  
}
\label{fig2} 
\end{center}
\end{figure}

We next investigate what occurs for the original CAMP with $\theta=0$.  
In Theorem~\ref{theorem3}, we have assumed the convergence of the asymptotic 
covariance~$d_{\tau,t}$ given in (\ref{Gaussianity}). As shown in 
Fig.~\ref{fig2}, a soliton-like wave propagates for $\theta=0$ while 
the convergence assumption in Theorem~\ref{theorem3} is valid for 
$\theta=-0.7$. 

The reason why the soliton-like wave occurs is as follows: 
As the condition number $\kappa$ increases, $a_{\tau,t}$ in the SE 
equation~(\ref{SE_equation}) becomes unstable. On the other hand, the 
forgetting factor $\bar{\xi}_{t}\in(0,1]$ in (\ref{xi_bar}) decreases 
in general as $a_{t,t}$ grows. As a result, a soliton-like quasi-steady wave 
appears when $a_{\tau,t}$ is unstable. It is an interesting future issue to 
analyze the stability of the SE equation~(\ref{SE_equation}).     

We finally present numerical simulations for the CAMP, AMP~\cite{Donoho09}, 
and OAMP/VAMP~\cite{Ma17,Rangan192}. See Table~\ref{table1} for the 
complexity of these algorithms. As long as the number of iterations $t$ 
is sufficiently smaller than $M$, CAMP has the same complexity as AMP, while 
OAMP/VAMP requires higher complexity.

\begin{table}[t]
\begin{center}
\caption{
Complexity in $M\leq N$ and the number of iterations~$t$. 
}
\label{table1}
\begin{tabular}{|c|c|c|}
\hline
& Time complexity & Space complexity \\
\hline
CAMP & ${\cal O}(tMN + t^{2}M + t^{4})$ & ${\cal O}(MN + tM + t^{2})$ \\
\hline
AMP & ${\cal O}(tMN)$ & ${\cal O}(MN)$ \\
\hline
OAMP/VAMP &  ${\cal O}(M^{2}N+tMN)$ & ${\cal O}(N^{2}+MN)$ \\
\hline
\end{tabular}
\end{center}
\end{table}

We used a damping technique to improve the convergence properties of the three  
algorithms. See \cite[Sec.~III-E and Sec.~IV-A]{Takeuchi203} for the details 
of damping. The parameter $\theta$ and damping factor were optimized via 
exhaustive search.

\begin{figure}[t]
\begin{center}
\includegraphics[width=\hsize]{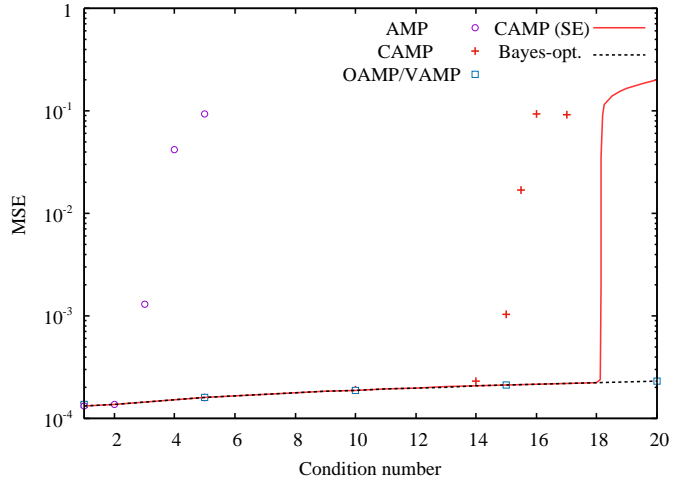}
\caption{
MSE versus the condition number~$\kappa$ for the CAMP, AMP, and OAMP/VAMP. 
$M=2^{10}$, $N=2^{11}$, $\rho=0.1$, $1/\sigma^{2}=30$~dB, $100$ iterations, and 
$10^{5}$ independent trials.  
}
\label{fig3} 
\end{center}
\end{figure}

As shown in Fig.~\ref{fig3}, AMP approaches the Bayes-optimal MSE only for 
small condition numbers. On the other hand, the CAMP is Bayes-optimal for  
low-to-moderate condition numbers. However, the performance of CAMP degrades 
for high condition numbers, for which OAMP/VAMP still achieves the 
Bayes-optimal MSE. These results imply that CAMP has room for improvement 
in the case of high condition numbers while it can improve the convergence 
property of AMP. 

\section*{Acknowledgment}
The author was in part supported by the Grant-in-Aid 
for Scientific Research~(B) (JSPS KAKENHI Grant Numbers 18H01441 and 
21H01326), Japan. 



\balance
\bibliographystyle{IEEEtran}
\bibliography{IEEEabrv,kt-isit2021}


\end{document}